\documentclass[a4paper,11pt]{article}

\pdfoutput=1 

\usepackage{jinstpub} 
\usepackage{subcaption}
\usepackage{float}
\usepackage{eurosym}

\title{\boldmath First demonstration of 3D optical readout of a TPC using a single photon sensitive Timepix3 based camera}

\author[a,1]{A. Roberts,\note{Corresponding author.}}
\author[b,c]{P. Svihra,}
\author[d]{A. Al-Refaie,}
\author[e]{H. Graafsma,}
\author[d,f]{J. K\"{u}pper,}
\author[a]{K. Majumdar,}
\author[a]{K. Mavrokoridis,}
\author[g]{A. Nomerotski,}
\author[e]{D. Pennicard,}
\author[a]{B. Philippou,}
\author[d]{S. Trippel,}
\author[a]{C. Touramanis,}
\author[a]{J. Vann}

\affiliation[a]{University of Liverpool, Department of Physics, Oliver Lodge Bld, Oxford Street, Liverpool, L69 7ZE, UK}

\affiliation[b]{Department of Physics, Faculty of Nuclear Sciences and Physical Engineering, Czech Technical University, Prague 115 19, Czech Republic}

\affiliation[c]{School of Physics and Astronomy, The University of Manchester, Manchester M139PL, United Kingdom}

\affiliation[d]{Center for Free-Electron Laser Science, Deutsches Elektronen-Synchrotron DESY, Notkestra{\ss}e 85, 22607 Hamburg, Germany}

\affiliation[e]{Deutsches Elektronen-Synchrotron DESY, Notkestra{\ss}e 85, 22607 Hamburg, Germany}

\affiliation[f]{Department of Physics, Universit\"{a}t Hamburg, Luruper Chaussee 149, 22761 Hamburg, Germany}

\affiliation[g]{Brookhaven National Laboratory, Upton, NY 11973, U.S.A.}

\emailAdd{aroberts@hep.ph.liv.ac.uk}

\abstract{\\The ARIADNE project is developing innovative optical readout technologies for two-phase liquid Argon time projection chambers (LArTPCs). Optical readout presents an exciting alternative to the current paradigm of charge readout. Optical readout is simple, scalable and cost effective. This paper presents first demonstration of 3D optical readout of TPC, using CF$_{4}$ gas as a proof of principle. Both cosmic rays and an Americium-241 alpha source have been imaged in 100~mbar CF$_{4}$. A single-photon sensitive camera was developed by combining a Timepix3 (TPX3) based camera with an image intensifier. When a pixel of TPX3 is hit, a packet containing all information about the hit is produced. This packet contains the $x,y$ coordinates of the pixel, time of arrival (ToA) and time over threshold (ToT) information. The $z$ position of the hit in the TPC is determined by combining drift velocity with ToA information. 3D event reconstruction is performed by combining the pixel's $x,y$ location with this calculated $z$ position. Calorimetry is performed using time over threshold, a measure of the intensity of the hit.}

\keywords{Time projection Chambers (TPC), Noble liquid detectors, Micropattern gaseous detectors, Photon detectors for UV, visible and IR photons (solid-state).}

\collaboration{\hspace{37mm} \includegraphics[height=30mm]{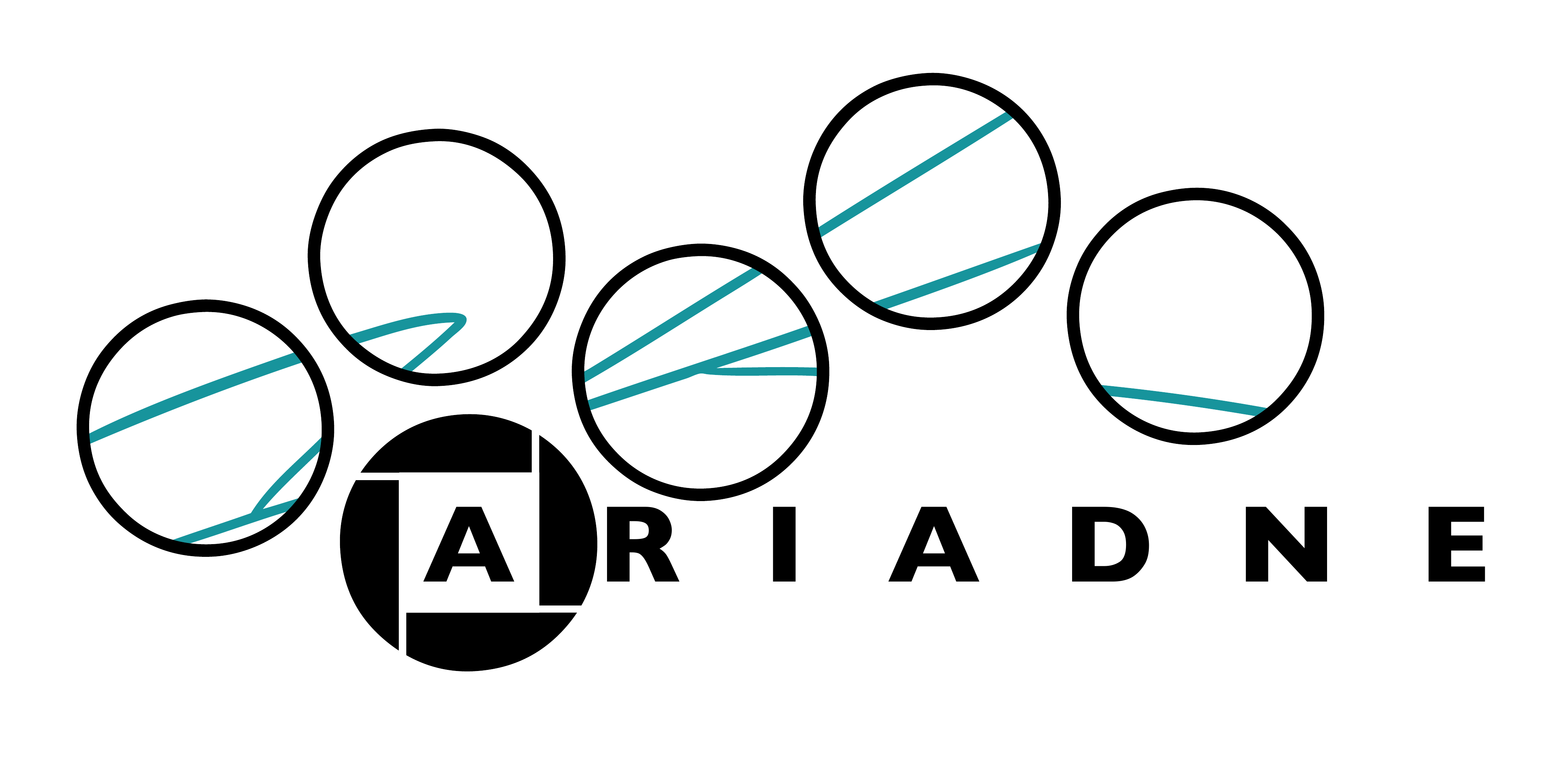}}

\begin{document}

\maketitle

\flushbottom

\section{Introduction and motivation}


Colossal LAr TPCs are considered by many to be the most promising far detector technology for future long baseline neutrino experiments. Realisation of these large-scale LAr TPCs is underway, with the DUNE collaboration bringing together single and two-phase LAr TPC technologies \cite{DuneCDRVol1, DuneCDRVol2, DuneCDRVol3, DuneCDRVol4, DuneIDRVol1, DuneIDRVol2, DuneIDRVol3}. Optical readout is a simple, scalable and cost effective methodology that presents an exciting alternative to the current paradigm of charge readout. The optical readout approach presented in this paper is compatible with both two-phase and gas TPCs. 

Thick gaseous electron multipliers (THGEMs) are commonly used for two-phase charge readout. THGEMs provide a large gain on the initial number of electrons that are ionised in the TPC. Compared to single-phase TPCs, this large charge gain can offer improved signal-to-noise ratios. Yet, THGEM charge readout is not without difficulties. To produce a large charge gain, a large potential difference must be applied across the THGEM, which can present long term stability issues. Traditionally, the charge produced in the THGEM is collected using a segmented anode. The number of readout channels can grow quickly with detector size. The nature of charge readout can introduce vulnerabilities to electronic noise. Care must be taken to avoid creating ground loops between different elements of the detector and also to sufficiently shield each readout channel.

Optical readout instead exploits the secondary scintillation light produced in the THGEM holes \cite{GEMLight1, GEMLight2, GEMLight3, GEMLight4, GEMLight5}. We have already demonstrated the imaging of cosmic muons with a 40 litre LAr ARIADNE prototype TPC utilising a THGEM and Electron Multiplying CCD (EMCCD) camera \cite{DemonstratorCCD, DemonstratorEMCCD}. ARIADNE is a research project for the development of innovative optical two-phase LArTPC readouts \cite{ARIADNETDR, ARIADNEProceedings, ARIADNELoI, ARIADNEWebsite}. This paper presents exciting progress that has been made in optical readout methodology. 3D tracking with zero charge readout channels is now possible.  

\section{Experimental setup}
 
\subsection{The ARIADNE prototype TPC}

The ARIADNE prototype detector is built around a cylindrical TPC. The field cage of the TPC is 20~cm long with 178~mm diameter. A cryogenic 8-inch Hamamatsu R5912-20 PMT, coated with Tetraphenyl butadiene (TPB), is installed below the TPC. Above the field cage, it is possible to install various devices such as THGEMs or MicroMegas \cite{micromegas} . Dual stacked THGEMs were installed above the TPC for the results presented in this paper. The THGEMs used in this work have typical dimensions, both 1~mm thick containing a hexagonal array of through holes. Each hole has 500~$\mu$m diameter and the hole-to-hole pitch is 800~$\mu$m. A 50~$\mu$m dielectric rim is etched around each hole to extend the breakdown voltage. Each THGEM has a 150 cm$^2$ octagonal active area. The optical transparency of the THGEM is roughly 35\%. Next to the TPC, suspended from a rotary feedthrough, is an Americium-241 alpha source. This alpha source may be rotated inside of the TPC to generate alpha tracks in the active volume. By rotating the alpha source outside of the TPC, only cosmic events are seen in the active volume. 

The camera (discussed further in section \ref{TPX3CamSection}) is mounted on top of the detector, outside of the TPC volume. The camera looks through a 90~mm diameter glass viewport, at a distance of 1~m from the THGEMs.  The transparency of the glass viewport is better than 90\% for the scintillation wavelength of CF$_{4}$ (620~nm). 

 \subsection{TPX3Cam} \label{TPX3CamSection}

TPX3Cam \cite{TPX3Cam, TimepixCam, TPXCamCharacterization}  uses a light sensitive silicon sensor bump bonded onto a TPX3 chip \cite{TPX3Chip}. It has an array of 256 x 256 pixels of 55~$\mu$m size, with each pixel in the sensor being connected to its own set of signal processing circuitry in the chip. When photons are absorbed by a pixel of the silicon sensor, electron-hole pairs are excited. The circuitry in Timepix3 accumulates these electrons in a capacitor, and if the accumulated charge exceeds a certain threshold (defined by the user) then the pixel registers a hit. Using an internal clock within the chip, the pixel records the time at which the hit occurred, i.e. the Time of Arrival (ToA). This timestamp has 1.6 ns resolution. Then, the capacitor is discharged at a constant rate, and the pixel records the length of time for which the charge in the capacitor remains above the threshold, i.e. the Time over Threshold (ToT). This value is proportional to the number of photons absorbed by the pixel. Finally, the pixel sends out a packet of data containing its $x, y$ location, the Time of Arrival and the Time over Threshold. This packet is referred to as a hit.

TPX3 has a minimum operating threshold of about 500$e^-$, required to overcome the electronic noise of the chip. The direct photon flux from the THGEM, which is between ten to a few hundred photons per pixel, is not high enough to directly overcome this threshold. An image intensifier is required in front of the camera to boost the intensity of the light signal. The 18~mm diameter intensifier boosts the incident light signal using a multi-stage process. First, the incoming photons are converted to electrons using a photocathode. These electrons are multiplied using microchannel plates (MCPs), with a gain of up to 1E6. Finally, these multiplied electrons are converted back to light using a scintillating screen. For these measurements, a Photonis cricket \cite{PhotonisCricket} image intensifier was used. The quantum efficiency of the Hi-QE-red photocathode \cite{Photonis} of the intensifier is $\approx$ 20$\%$ at 620~nm (The peak scintillation wavelength of CF$_{4}$). The dark count rate of the photocathode is between 30-60~kHz. The intensifier uses two MCPs, arranged in a chevron configuration, giving a total electron gain of 85 000. The intensifier uses a fast scintillating screen, P47, with a rise time of 7~ns \cite{P47Risetime}.  

The non-intensified version of TPX3Cam has been used for ion and electron imaging \cite{tpx3camVMI}. The intensified version of TPX3Cam has been used for spatial characterization of photonic polarization entanglement \cite{tpx3camQIS}. It has also been used for photon counting phosphorescence lifetime imaging \cite{tpxcamFLIM}.

\section{Detection principle} 

When a charged particle passes through the TPC, prompt scintillation light is produced and the gas CF$_{4}$ is ionised. The liberated electrons drift towards the THGEMs. When electrons enter the holes of the THGEM, they experience the high electric field applied across the THGEM. In large fields, the electrons gain enough kinetic energy to ionise gas molecules. This releases more electrons, which can in turn go on to further ionise the gas. The resulting electron avalanche is known as a Townsend discharge. The charge cloud produced in the THGEM hole also excites many gas molecules. As these gas molecules de-excite, large amounts of secondary scintillation light are produced. 

\begin{figure}

\begin{center}
\includegraphics[width=0.92\textwidth]{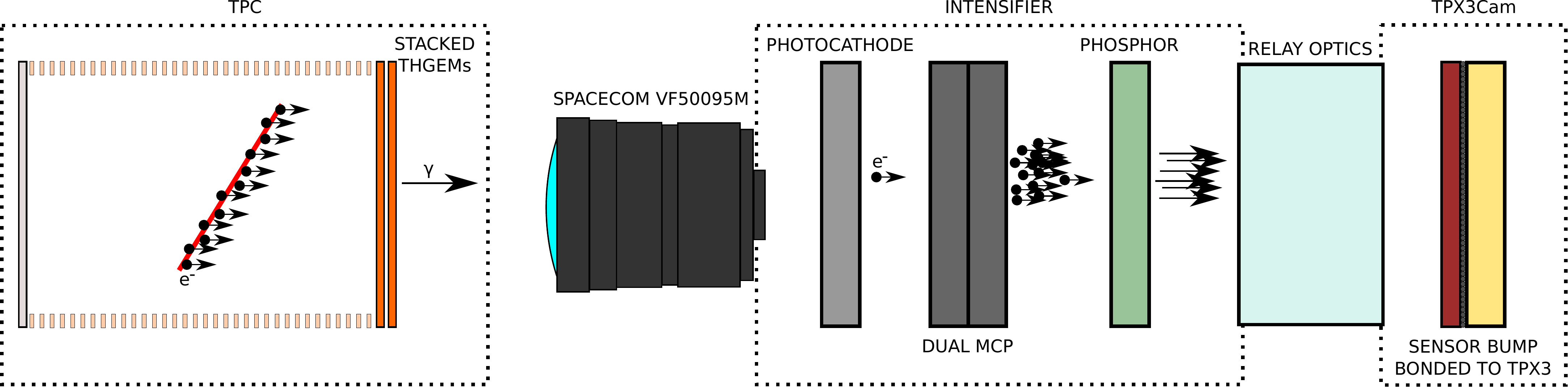}

\caption{Detection principle. Secondary scintillation light is produced during the charge multiplication process in the THGEM holes. This light is detected by the externally mounted camera. An image intensifier boosts the incident light signal. The amplified light signal is detected by TPX3Cam.}

\label{detectionPrinciple}
\end{center}
\end{figure}

The secondary scintillation light leaves the detector volume through a glass viewport and is captured by the external intensified TPX3Cam. 3D tracking is performed by combining the hit pixel's $x,y$ location with simultaneous time of arrival information. $x,y$ position of the track is found by calibrating the field of view of the camera to the known THGEM dimensions. $z$ position of the hit in the TPC is calculated by combining ToA information with the known drift velocity in the TPC. Calorimetry is performed using ToT, which is a measure of hit intensity. The detection principle is shown in Figure \ref{detectionPrinciple}.

\begin{figure}
\begin{center}
\includegraphics[width=0.92\textwidth]{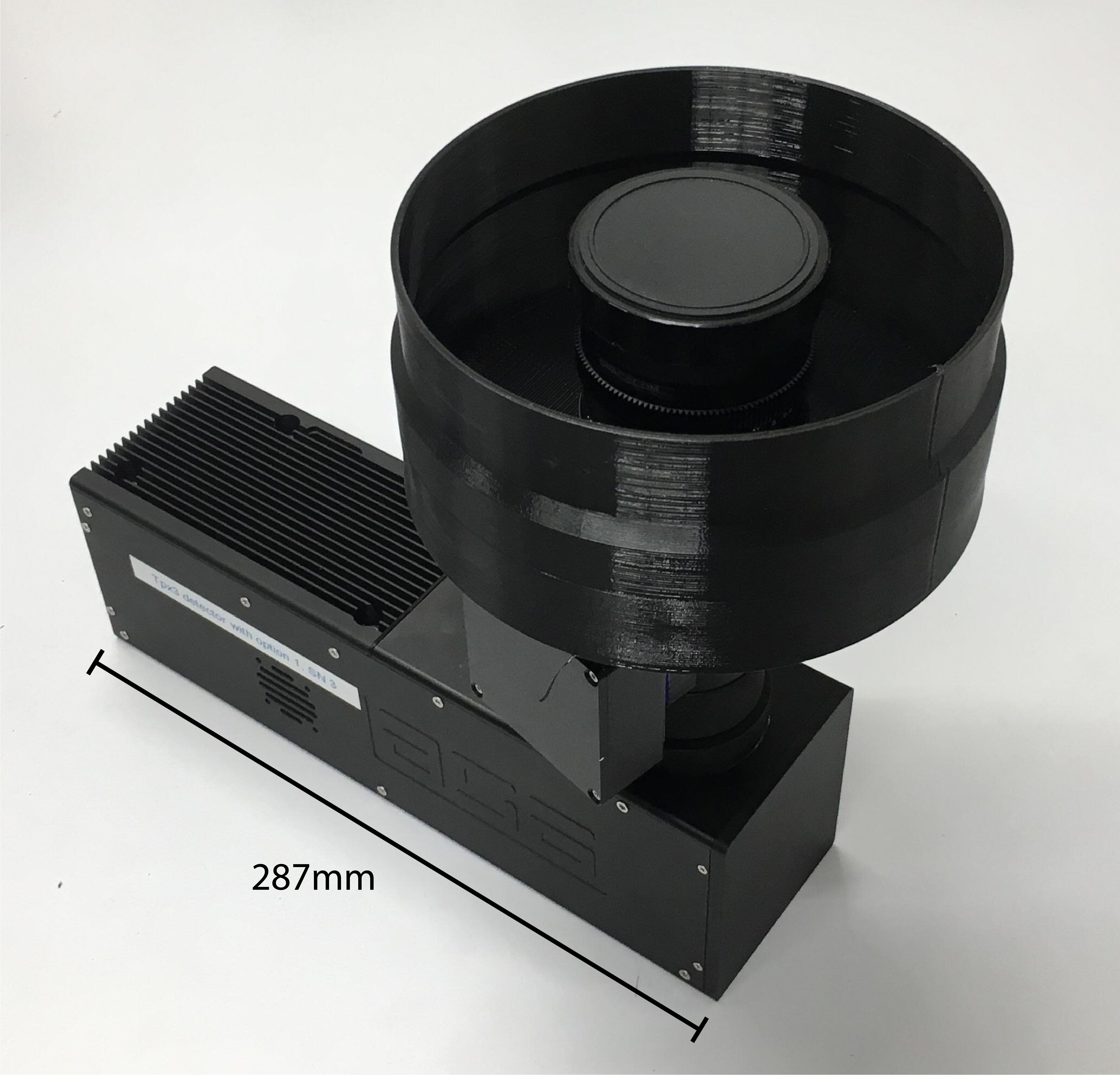}

\caption{Intensified TPX3Cam. 3D printed bellows are used to create a light tight seal between the detector and camera. The lens has a protective cap installed, protecting the photocathode of the intensifier.}

\label{TPXCamSetup}
\end{center}
\end{figure}

\begin{figure}
\begin{center}


\begin{subfigure}{1.\textwidth}
\centering
\includegraphics[height=180mm]{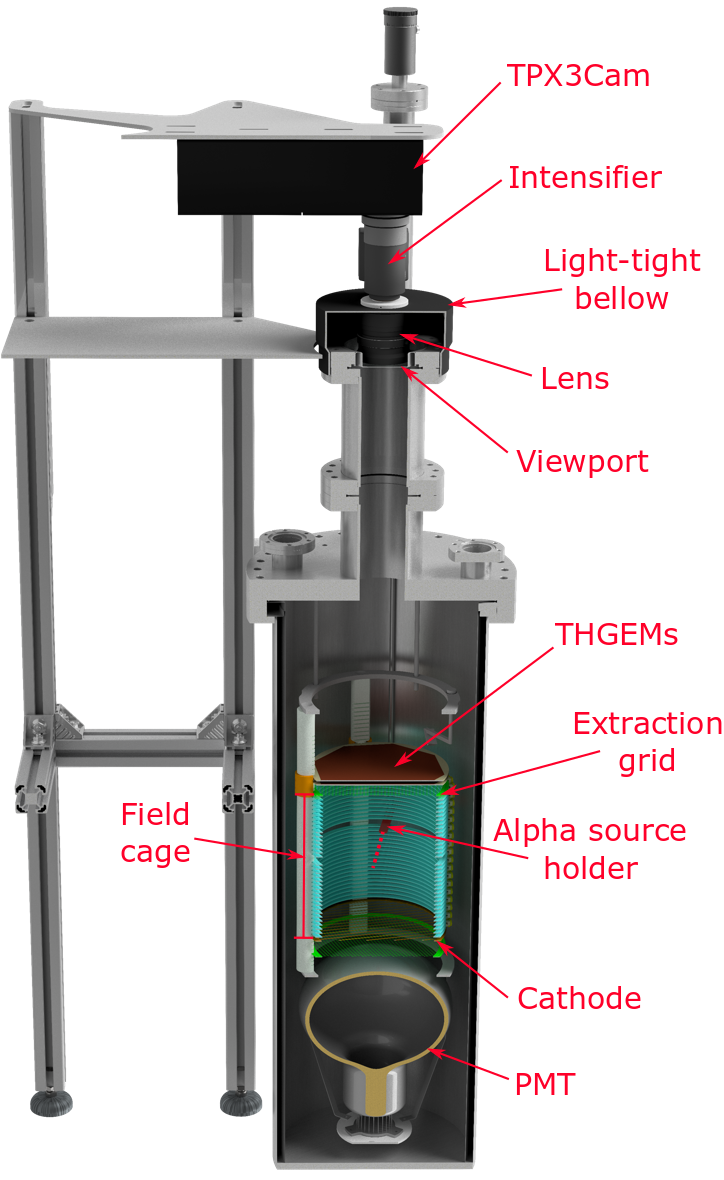}
\end{subfigure}%
\caption{Experimental setup. The intensified TPX3Cam looks through a glass viewport towards the THGEMs. A Spacecom VF50095M lens captures the light that is produced in the THGEM holes.}

\label{ExperimentSetup}
\end{center}
\end{figure}

A Spacecom VF50095M lens is installed in front of the image intensifier. Chosen to maximise light collection, this lens has a speed of f/0.95. The 50~mm focal length of the lens gives the camera a field of view covering roughly 18~cm $\times$ 18~cm. This field of view yields an $x,y$ resolution of 0.7~mm/pixel. Given the drift velocity of $\approx$ 10~cm/$\mu$s in the TPC, the 1.6~ns ToA resolution gives a $z$ position resolution better than 1~mm. As demonstrated by the measurements in this paper, the ToA resolution of TPX3 allows for high resolution 3D tracking, even in gas TPCs with fast drift velocities. When considering the slower drift velocities of noble liquid TPCs, the $z$ position resolution is expected to be diffusion limited. The $x,y$ resolution of optical readout is flexible according to lens selection. High $x,y$ resolution is possible using lenses with a small field of view. Wider fields of view trade $x,y$ resolution for cost savings by requiring less cameras to cover the same area. Figure \ref{TPXCamSetup} shows the camera and Figure \ref{ExperimentSetup} shows the camera installed on top of the detector.

\newpage

\section{Results} \label{resultsSection}

The TPC was filled with 100~mbar CF$_{4}$ gas for all measurements. The cathode was biased to -3000~V and the extraction grid to -1000~V, giving a drift field of 100~V/cm. The bottom of the first THGEM was connected to negative polarity bias voltage. The top of the first THGEM and the bottom of the second THGEM were both grounded. The top of the second THGEM was connected to positive polarity bias voltage.

The threshold of TPX3 was set to about 1000~$e^-$. This value is so far above the noise floor of TPX3 that when the intensifier was switched off, no hits are seen. In practice, the exact threshold value was not found to be very important. Due to the large gain of the intensifier, a single incident photon results in tens of thousands of photons reaching TPX3Cam. The threshold simply should be set above the noise floor of TPX3 but below the order 10,000~$e^-$ signal for a single incident photon at the intensifier. 

For arbitrary tracks, the absolute $z$ position of a hit may be determined by combining camera measurements with PMT measurements. The PMT will detect primary scintillation light with good efficiency. The time delay between primary and secondary scintillation light determines the depth of the track in the TPC. In this work however, primary scintillation light was not seen by the camera. This is probably due to the low optical transparency of the THGEM coupled with the already low number of photons produced by primary scintillation compared to secondary scintillation. This work instead presents relative track information in $z$. The topmost hit of the track is defined as $z=0$ for each event. Future work will investigate correlation of PMT measurements with camera measurements. 

\begin{figure}
\begin{center}
\includegraphics[width=1.\textwidth]{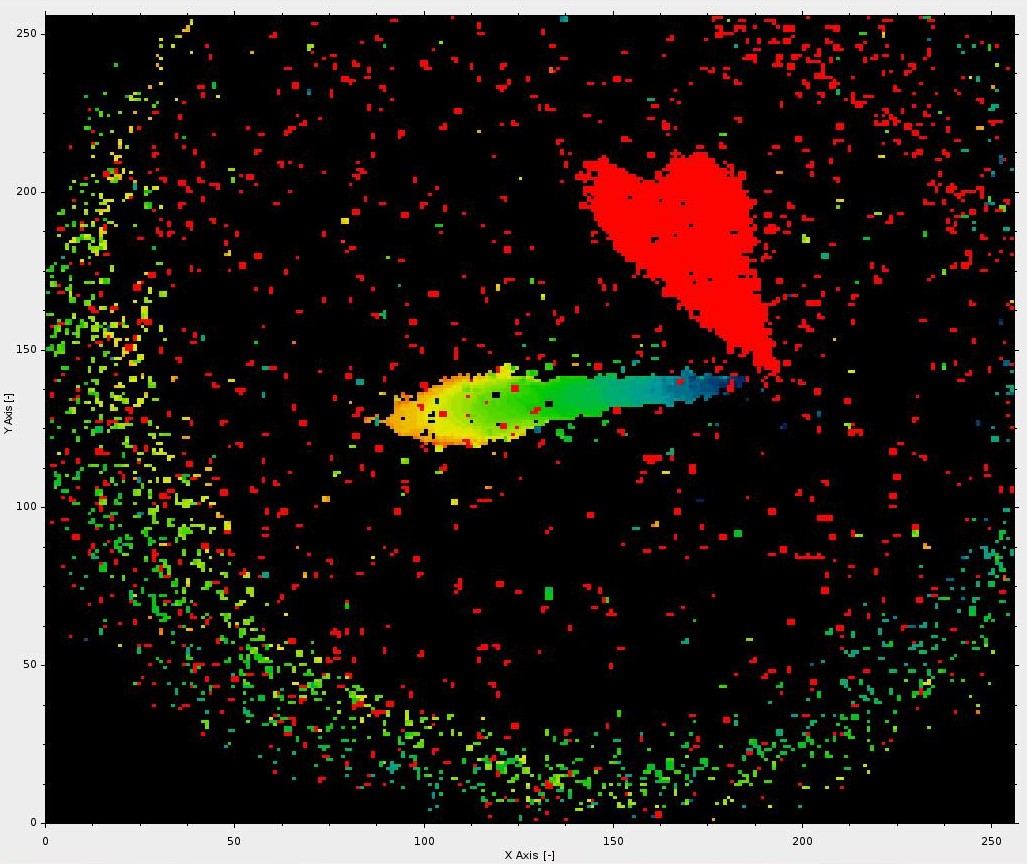}

\caption{Simulated 1~ms exposure captured using SoPhy. Colour shows ToA. One alpha track is seen to have a changing colour scale over the length of the track. This changing colour scale implies that the ionised track arrived at the THGEM over time. If a track was horizontal in the TPC, all charge from the track would arrive at the THGEM at the same time. Thus, we can imply that this alpha track is at an angle relative to horizontal. The exact value of this angle is determined by the rate of change of the colour scale. Reflections from the viewport are visible around the edge of the image. These reflections are seen to arrive at the same time as the light from the alpha track. Two other tracks are visible on the image, in the overflow bin (red colour) of the ToA colour scale. Each THGEM had a bias of 780~V for this measurement.}

\label{SoPHYToA}
\end{center}
\end{figure}

Data from TPX3Cam can be visualised in two different ways. First, it is possible to view the data from the camera as a 'photograph' with a simulated exposure time. By integrating all hits received during a specified time window, a 2D image is produced. SoPhy software, provided with TPX3Cam, provides live video output using this approach. Certain diagnostic tasks, such as lens focusing, are much easier to perform using a live video stream. Figure \ref{SoPHYToA} shows several alpha tracks imaged using SoPhy.

The native operating mode of TPX3 is data driven. In this mode, each of the 256 $\times$ 256 pixels operates independently. If any pixel detects a signal over the preset threshold, the pixel's $x,y$ position plus ToT and ToA information are recorded. This data is sent off camera asynchronously via an Ethernet or Fiber connection. Each pixel operating independently allows for sparse readout of the sensor. Any event in the TPC which produces light above the preset threshold will be read out. Regions of the TPC which produce light below the threshold will be zero suppressed. As long as pixel occupancy is modest, much higher effective frame rates are possible compared to full frame readout. In colossal TPCs, the dimensions of events are expected to be small compared to the total readout area. Thus, data driven readout is a very efficient approach for reading out large area TPCs. Data driven readout of a TPC is continuous; no dedicated hardware trigger is required. However, if required, TPX3 is still able to accept external input signals. When an external signal is received, a time-stamp is inserted into the data stream from the camera. This time-stamp marks the time at which the external signal was received with a resolution of 260~ps. Specific events, marked by this external time-stamp, can be easily be identified.  
 
\begin{figure}
\begin{center}
\includegraphics[width=1.\textwidth]{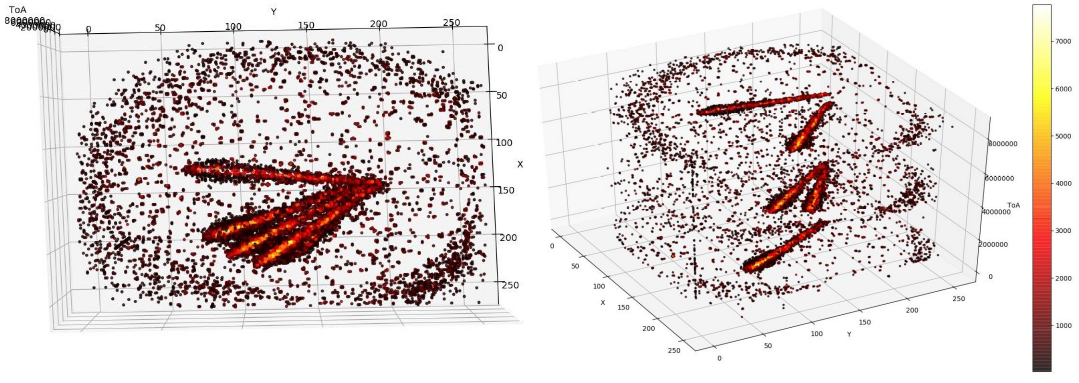}
\caption{Raw data captured by the intensified TPX3Cam for an 8ms time window. Left: Data arranged to simulate a traditional CCD image. Right: Data rotated to show the extra  dimension given by simultaneous ToA. Viewport reflections are seen as clusters of pixel hits around the edges of the image. These hits arrive simultaneously with the direct light from the alpha tracks. The randomly arriving low intensity pixel hits are explained by photocathode dark count. Each THGEM has a bias of 780~V. Continuous TPC readout was performed for 8~ms so the $z$ axis of these visualisations contain many detector volumes.}
\label{TraditionalComparison}
\end{center}
\end{figure}

\begin{figure}[H]
\begin{center}
\includegraphics[width=1.\textwidth]{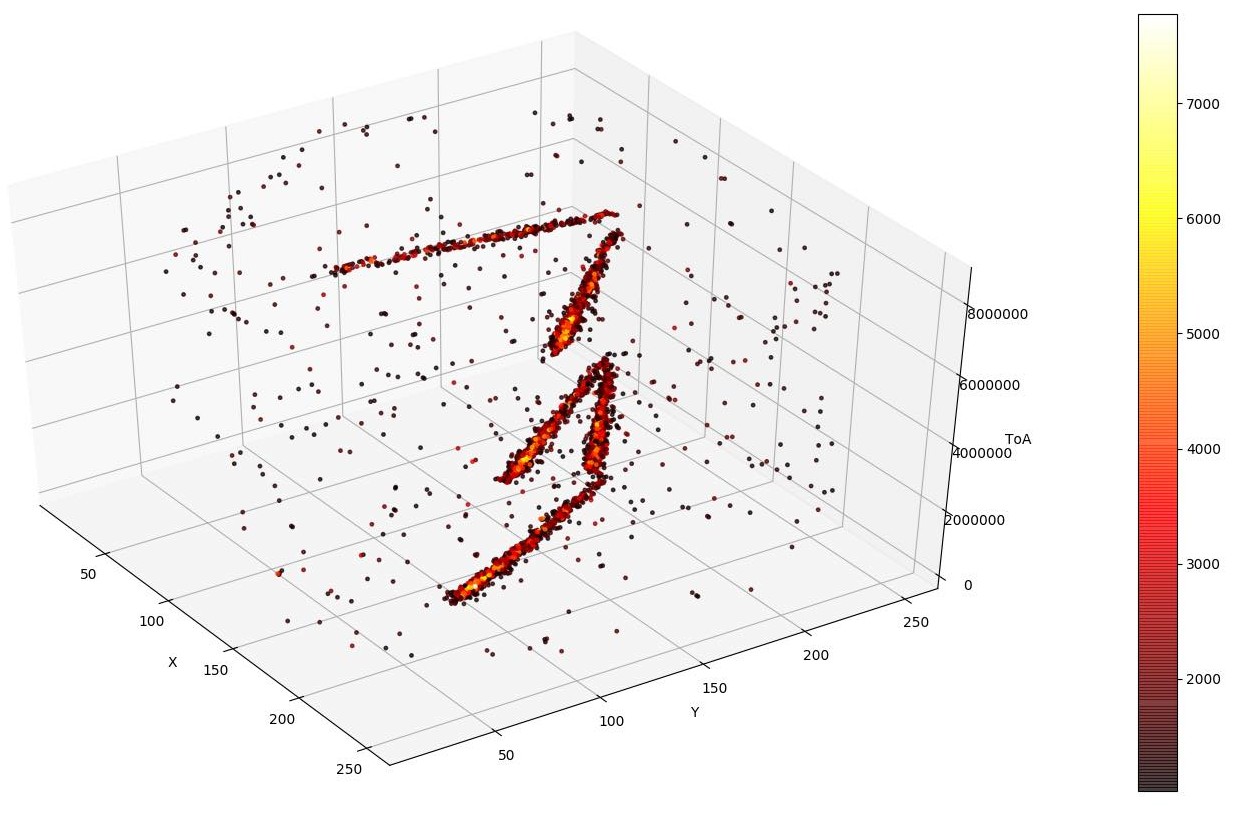}
\caption{3D visualisation of an 8~ms time slice. Colour represents ToT. A cut of ToT > 1000ns has been performed to remove many dark count hits plus viewport reflections. Five alpha tracks are visible during this 8~ms time slice. This rate is in agreement with the roughly collimated source rate of 1 kHz. A Bragg peak is visible at the end of the alpha tracks. Each THGEM has a bias of 780~V. 8~ms continuous TPC readout means that $z$ axis of this visualisation contains many detector volumes.}
\label{MultipleAlphas}
\end{center}
\end{figure}

Figure \ref{TraditionalComparison} illustrates the crucial detail that separates TPX3Cam from traditional cameras. On the left hand side of Figure \ref{TraditionalComparison} we have visualised the data to simulate a fixed 8ms exposure. This is how the image from a traditional CCD camera might look. Many alpha tracks can be seen, each overlapping with one other. Rotating the visualisation unveils the extra time dimension gained by simultaneous ToA information. ToA time-stamping allows each alpha track to be distinctly resolved in time. These visualisations plot raw hits with no cuts applied. Figure \ref{MultipleAlphas} plots these hits with a simple threshold cut of ToT > 1000~ns. This simple cut is already enough to produce very clean events. A Bragg peak is visible at the end of each track. 

In these visualisations we can also see several randomly positioned low intensity hits. These hits are a result of photocathode dark count. The Hi-QE red photocathode of the intensifier spontaneously emits photo-electrons at a rate of about 30-60kHz. This single photo-electron is amplified by the intensifier and is seen by the camera as a point-like hit with low intensity. Large clusters of hits can also be seen on the outermost edges of the visualizations. These hits are reflections from the stainless steel tube at the top of the detector. These hits are seen to arrive correlated in time with the light from the alpha tracks, consistent with being reflections. A hot pixel is visible at approximately (30,180). Each pixel of TPX3 has a deadtime of 475~ns + ToT. The hit rate that is allowed by this deadtime is much higher than the dark count rate of the photocathode. As a result, this pixel looks like it is receiving hits much more frequently in time than it's neighbours. 

\begin{figure}[H]
\begin{center}
\includegraphics[width=.6\textwidth]{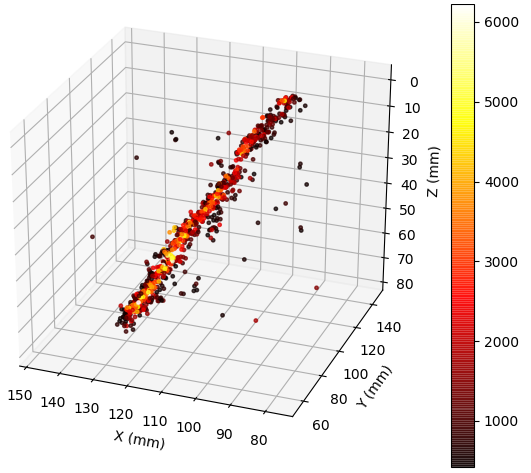}
\caption{3D reconstruction of a single alpha track. Colour represents ToT. Photocathode dark count is seen as the few low intensity pixel hits arriving randomly in time and position. ToA was converted to millimeters using the known drift velocity of 10~cm/$\mu$s in the TPC. $x,y$ pixel values were converted to millimeters by calibrating the field of view of the camera to the known THGEM dimensions. Each THGEM has a bias of 780~V.}
\label{SingleAlpha}
\end{center}
\end{figure}

Figure \ref{SingleAlpha} shows an individual alpha track. Conversion of the data to physical units is possible using simple multiplicative factors on each axis. The $x,y$ axes are converted to millimeters by calibrating the field of view of the camera to the known THGEM dimensions. The $z$ dimension of this plot has been converted to millimeters using the observed drift velocity of 10~cm/$\mu$s in the TPC.

\begin{figure}[H]
\begin{center}
\includegraphics[width=.62\textwidth]{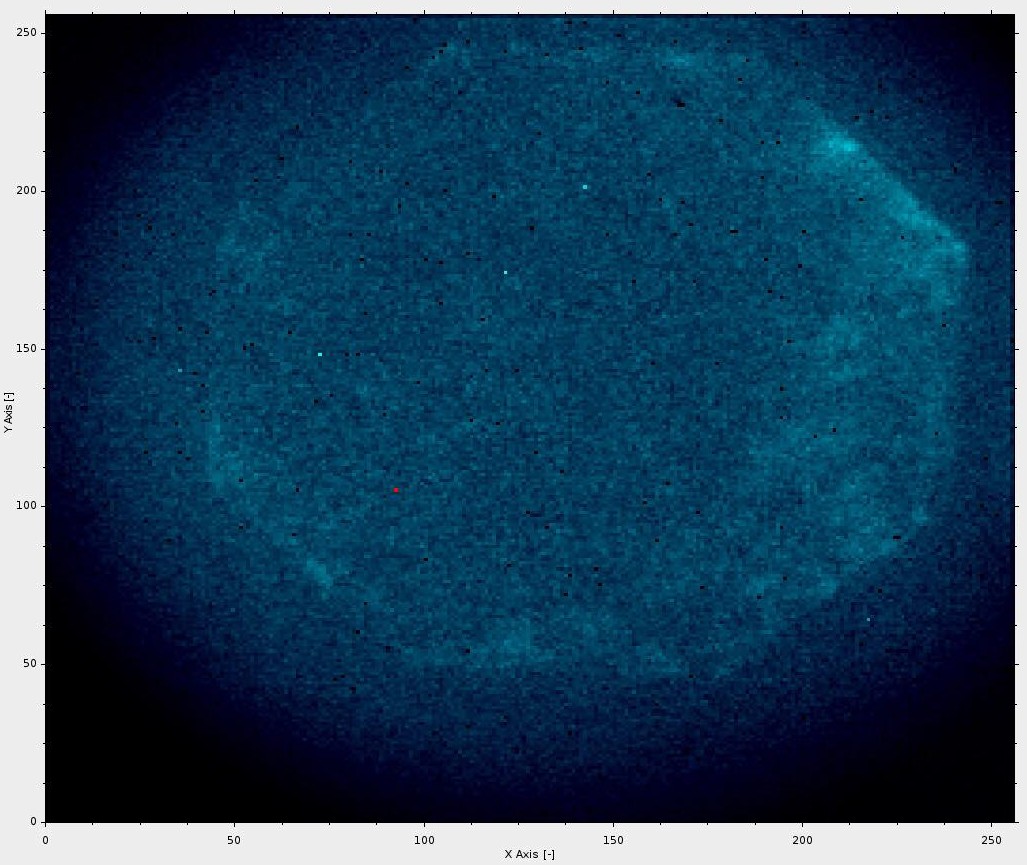}
\caption{10 second exposure of the THGEMs with alpha source removed. Colour represents ToT. The octagonal THGEM active area is illuminated by cosmics passing through the TPC. The THGEMs were biased to 860V for this demonstration.}
\label{GemIlluminated}
\end{center}
\end{figure}

\begin{figure}[H]
\begin{center}
\includegraphics[width=1.\textwidth]{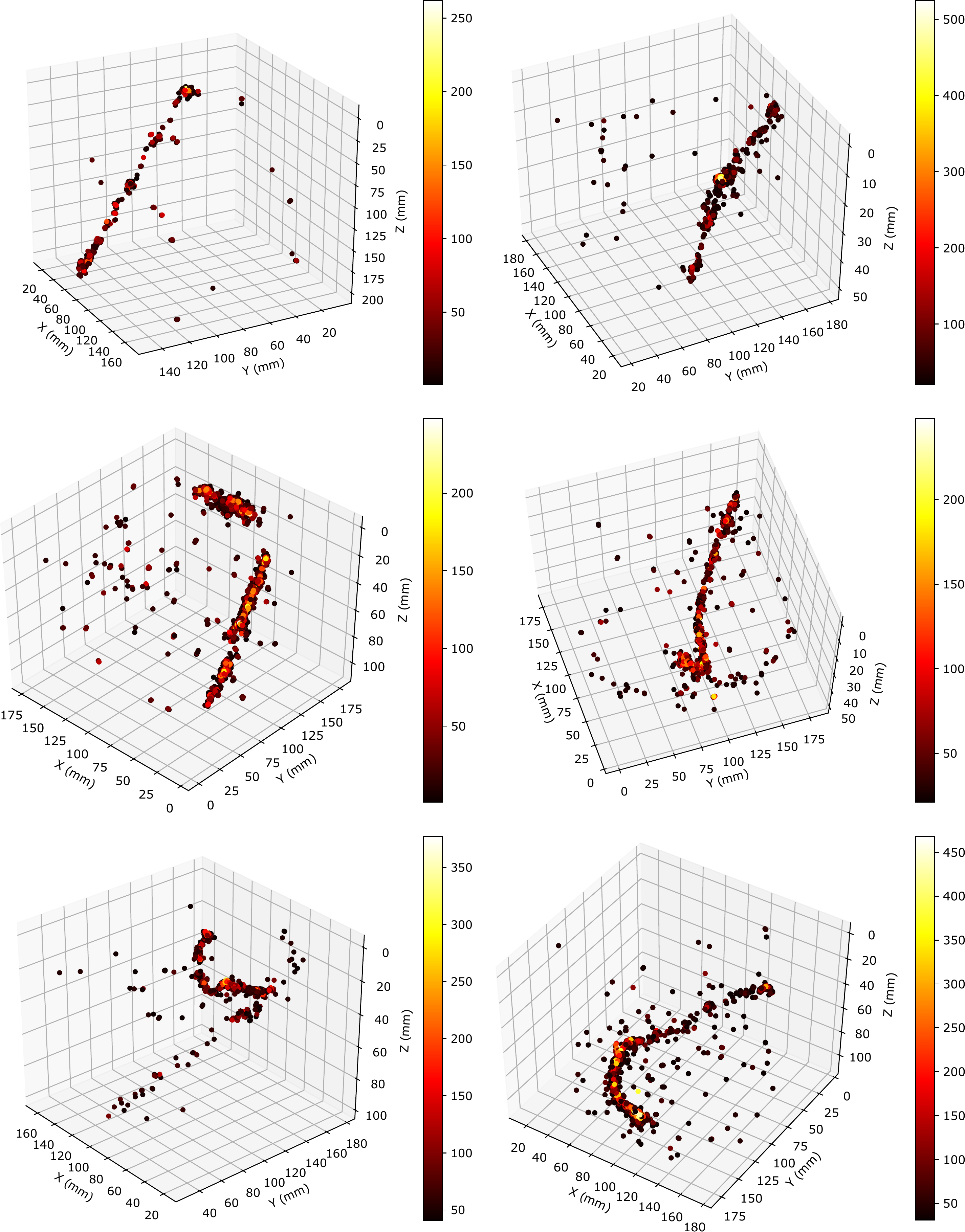}
\caption{A sample of cosmic events observed in the TPC. The low energy deposition of cosmics required that the THGEMs were biased to 860V for these measurements.}
\label{Cosmics}
\end{center}
\end{figure}

Figure \ref{GemIlluminated} shows a long exposure image of the THGEM with alpha source removed. Cosmic rays passing through the TPC illuminate the THGEM's octagonal active area. The dark corners of the image are the result of poor mapping of the intensifier's $\phi$18~mm photocathode onto the 14~mm $\times$ 14~mm TPX3 sensor. Figure \ref{Cosmics} shows a gallery of cosmic events in the TPC. Due to the lower energy deposition of cosmic muons, the THGEM bias was increased to 860V for these measurements. 

\begin{figure}[H]
\begin{center}
\includegraphics[width=0.63\textwidth]{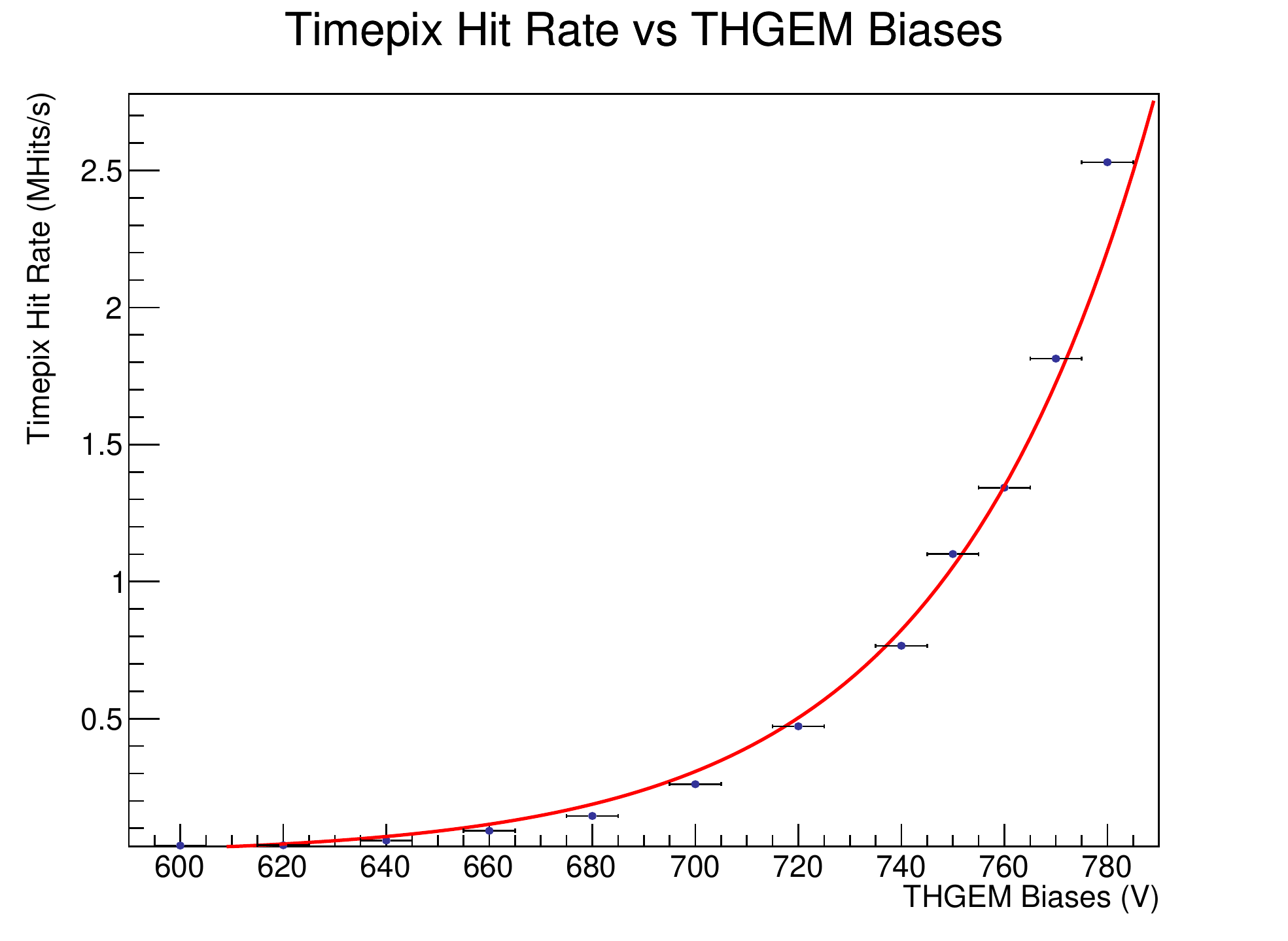}
\caption{Camera pixel hit rate vs applied THGEM bias. The bottom plane of the first THGEM was connected to negative polarity bias voltage. The top plane of the first THGEM and the bottom plane of the second THGEM were both grounded. The top plane of the second THGEM was connected to positive polarity bias voltage. The expected exponential increase in light production with increasing THGEM bias is well described.}
\label{THGEMBias}
\end{center}
\end{figure}

The response of the camera to varying THGEM bias was investigated. As the voltage across the THGEM is increased, charge gain and light production increase exponentially. At low THGEM bias, the alpha tracks are sparse. As light production increases, so does the density of the hits in the alpha tracks. Thus, the number of hits can be considered as a measure of light production. For this study, light production was measured using the hit rate from the camera. Figure \ref{THGEMBias} shows the hit rate of the camera versus bias across the THGEMs. The expected exponential increase in light production with THGEM bias is clearly shown.

\begin{figure}[H]
\begin{center}
\includegraphics[width=0.63\textwidth]{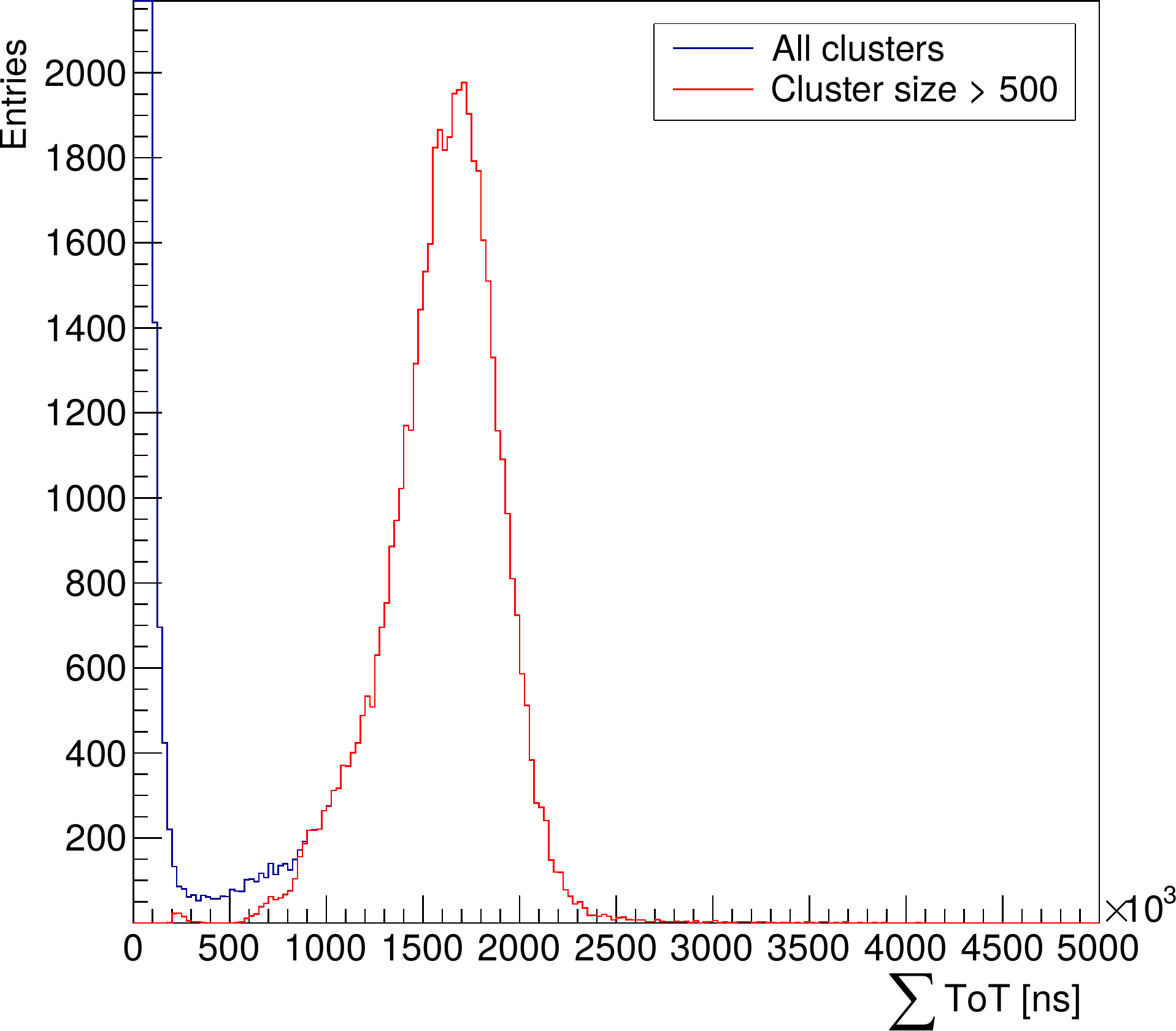}
\caption{Histogram of sum ToT values for all hits for a sample of alpha tracks. Hits were clustered together using the clustering algorithm described in more detail in Section \ref{resultsSection}. Cutting on cluster size > 500 provides a clean selection of alpha tracks. A peak is visible corresponding to the alpha energy of $\approx$ 5.5 MeV.}
\label{AlphaSpectrum}
\end{center}
\end{figure}

Calorimetric performance was also studied. ToT is a measure of the number of photons detected in each hit. This light intensity is proportional to the energy loss of an ionising particle in the TPC. For a contained track, the sum of ToT for all hits gives a measure of the total energy of the particle. Figure \ref{AlphaSpectrum} shows a distribution of summed ToT for alpha tracks in the TPC. Individual hits were clustered together and a cut was placed to select clusters containing more than 500 hits. The clustering algorithm looks for hits that are within a certain distance in the $x$, $y$ and $z$ directions of a chosen starting hit. The process is carried out recursively until all hits for a specific cluster have been found. The algorithm is then performed on any remaining hits not in the cluster to find new clusters. The cluster size cut of 500 hits was found to provide clean selection of alpha tracks over the sparse background hits. A clear peak can be seen corresponding to the alpha energy of 5.5~MeV. 

\section{Outlook and future developments}

\subsection{Performance}
 
Electroluminesence gain is much higher than charge gain in a THGEM; many more photons than electrons are produced per initial electron that enters in the THGEM. Electroluminescence yield versus charge yield has been studied for a 0.4mm thick THGEM in Argon gas. At a potential difference of 1200~V, the light yield was up to 1.5E4 photons per primary electron \cite{LightProductionThesis}. A bias of 1200~V across a 0.4mm thick THGEM translates to an electric field of 30~kV/cm. For this electric field in Argon gas, a charge gain of roughly 40 is expected \cite{THGEMChargeGain}. For absolute best detection thresholds, an array of photodetectors could be installed directly above the THGEM. However, there are other ways to exploit this large light yield. Since so much light is produced, it is possible to install the camera at a distance away from the THGEM. In this work, the 60mm diameter lens, roughly 1~m from the THGEM, gives a solid angle of 0.003 sr. The lens is expected to collect about 0.02 percent of all light produced. Photon collection efficiency has been traded for a reduction in the number of photodetectors. This may translate to large reductions in cost. Additionally, since the camera is now external, easy access is also realised. This is helpful for maintenance and future upgradability.

Despite the reduced photon collection efficiency that results from moving the cameras away from the THGEM, good detection thresholds are still realised. For a THGEM field of 30kV/cm in Argon gas, about 1.5E4 photons \cite{LightProductionThesis} are produced for every electron that enters the THGEM. The 0.02 percent collection efficiency of this setup predicts that three photons will reach the lens. Given the single photon sensitivity of the camera, three photons may be a detectable signal. Without the intensifier, orders of magnitudes more photons would be required to overcome the roughly 1000 electron readout noise of TPX3. The extra gain stage provided by the image intensifier greatly improves detection thresholds. 

Possible sources of light noise are well understood. There are two main sources of noise; light leaks and photocathode dark count. In this work, a seal was created between the camera and the TPC using a 3D printed flexible bellow. No difference in hit rate was seen when shining a torch on this joint so it was deemed to be light-tight. It is believed that it is not particularly difficult to create a light-tight seal at this interface and eliminate this first source of light noise. The second source of noise, photocathode dark count, depends largely on image intensifier configuration. Image intensifiers  are  available with dark count rates as low as 25~cps/cm$^2$. In the case of a 18~mm diameter image intensifier, as used in this work, 25~cps/cm$^2$ translates to a typical dark count rate of approximately 60~Hz. If we consider a colossal LArTPC with a drift length of 12~m and a typical drift velocity of 0.16~cm/$\mu$s, we have a maximum event window of just over 7~milliseconds. A dark count rate of 60~Hz suggests an average of 0.4 dark count hits per event per camera. It is not anticipated that light noise sources will pose significant problems following proper image intensifier reconfiguration. 


\subsection{Future outlook and developments}

Future applications for optical readout will be explored. Optical readout of large scale LAr TPCs is of particular interest. Noble liquid TPCs have much slower drift velocities when compared to gas TPCs. This translates to expected diffusion limited $z$ position resolution. A summary of the merits of TPC optical readout is presented below;

\begin{itemize}

\item Native 3D data format. Traditional strip based charge readout requires combining several planes of data to perform 3D reconstruction. Optical readout requires no combination of readout planes. Complications or errors arising from this reconstruction stage are eliminated.

\item Zero suppression is a natural by product of TPX3 data driven readout. TPX3 provides efficient compression of raw data. Each pixel hit from the camera is fully defined using only 16~bytes. As an example, the alpha tracks presented in this paper each contain roughly 1000 hits. Thus, raw data for each track is only 16~kB in size.

\item Allows for high readout rates (80~Mhits/s/camera). Related to the previous point, TPX3 enables very high event rates. The maximum readout rate is constrained by the bandwidth of the 10 Gigabit Ethernet link from each camera. This Ethernet connection allows for a maximum readout rate of 80~Mhits/s/camera. Again we will use the alpha tracks in this paper as example, each composed of 1000 hits. Assuming reflections are eliminated and dark count rate can be reduced to $\approx$~50~Hz, an alpha source with a decay rate of up to 80~kHz could have been imaged.

\item Fast timing resolution. The timing resolution of TPX3, as demonstrated in this paper, is fast enough to readout gas TPCs. Long baseline neutrino experiments generally feature both a near detector and a far detector. Optical readout of both a gaseous near detector and a liquid Argon far detector is conceivable. Using the same readout technology for both near and far detectors has many advantages. Common reconstruction and analysis software can be developed to process data from both detectors. As an example, consider the case of comparing near and far detector flux. Any readout based biases that exist in the data would be common between both detectors, possibly leading to reduced detector based systematics.

\item Triggerless / Continuous readout. No dedicated hardware trigger is required for optical readout with TPX3 based cameras. Hardware triggers often can have different efficiencies for selecting different types of event. As a result, hardware triggers can often introduce biases into data. The self triggering nature of TPX3 makes it very difficult to introduce a trigger bias into raw data.

\item Flexibility. By tailoring the image intensifier configuration, it is possible to detect a wide range of wavelengths. In the case of this work, an image intensifier with a Hi-QE-red photocathode was used. This photocathode has a quantum efficiency of around 20$\%$ at 620~nm, the peak scintillation wavelength of CF$_{4}$. For optical readout of an Argon TPC, tetraphenyl butadiene (TPB) may be used to shift the 128~nm Argon scintillation light to 430~nm. Several photocathode options are available at 430~nm with good quantum efficiency. GaAsP photocathodes offered by Hamamatsu have 40$\%$ quantum efficiency and Hi-QE Green photocathodes from Photonis have 30$\%$ quantum efficiency. 

\item Optical readout is not vulnerable to TPC electronic noise sources. Cameras are mounted externally, far from the main sources of electronic noise in the TPC. The photons produced in the THGEM are immune to electronic noise as they travel from the TPC to the camera. It is not  required to pay special attention to the grounding and shielding of the camera.

\item Straightforward upgrade and maintenance. The cameras are mounted outside of the difficult to access cryogenic volume. This easy access allows the cameras to be easily maintained. Upgrade of the cameras during the lifetime of the experiment is straightforward. 

\end{itemize}


\noindent Future developments have the potential for further performance improvements. Timepix4 is aimed towards a smaller pixel pitch and improved time stamping capability \cite{TPX4Roadmap}. A higher resolution sensor could offer improved $x,y$ tracking resolution and improved ToT resolution could improve calorimetry performance. It has been shown that Timepix can be integrated with an image intensifier into a single package \cite{TPXPlanacon}. This integration eliminates the phosphor screen and massively reduces the overall size of the camera.


Immediate next steps include installing TPX3Cams onto the ARIADNE detector. Using a larger 50~cm $\times$ 50~cm THGEM, ARIADNE will allow a liquid Argon demonstration of this technology as well as a demonstration of scalability by tiling together multiple cameras.

\acknowledgments

The ARIADNE program is proudly supported by the European Research Council Grant No. 677927 and the University of Liverpool. This work was supported by the BNL LDRD grant 13-006. Supported by The Czech Ministry of Education, Youth and Sports (grant LM2015054).

\end{document}